\documentclass{ws-procs9x6}
\usepackage[colorlinks]{hyperref}

\begin{document}
\title{Detecting quark matter in the early universe \\by gravitational waves}
\author{S. Schettler$^*$, M. T. Boeckel and J. Schaffner-Bielich}
\address{Institute for Theoretical Physics, Philosophenweg 16, Heidelberg University, Germany\\
$^*$E-mail: s.schettler@thphys.uni-heidelberg.de}

\begin{abstract}
For large baryo\-chemical potential $\mu_\mathrm{B}$ strongly interacting matter might undergo a first order phase transition at temperatures $T \sim 100-200$ MeV. Within standard cosmology, however, $\mu_\mathrm{B}$ is assumed to be very small leading to a crossover. We discuss implications of a first order QCD transition at high $\mu_\mathrm{B}$ being consistent with current observations. In this contribution we concentrate on effects on the gravitational wave spectrum. There are other interesting cosmological signals as a modification of the power spectrum of dark matter, the production of stellar black holes, and the seeds for the extragalactic magnetic fields which we briefly address also.
\end{abstract}

\keywords{Cosmological QCD phase transition; gravitational waves.}
\bodymatter

%%%%%%%%%%%%%%%%%%%%%%%%%%%%%%%%%%%%%%%%%
\section{Introduction}
%%%%%%%%%%%%%%%%%%%%%%%%%%%%%%%%%%%%%%%%%

In the standard scenario of the hot big bang a period of inflation is assumed to have taken place in the very early universe. During inflation metric perturbations are created of which the tensorial part forms a spectrum of primordial gravitational waves (GWs) presumably propagating through spacetime until today. Not interacting with the medium they can retain information about the very early stages of the cosmos. Nevertheless, the evolution of the universe can leave imprints in the primordial GW signal because the GW spectrum depends on the behavior of the Hubble parameter $H$ given by the Friedmann equation
\[ H^2 \equiv \left( \frac{\dot a}{a} \right)^2 = \frac{8\pi G}{3} \rho \propto g(T) T^4. \]
%%%%%%%%%%
%%%%%%%%%%%%%%%%%%%%%%%%%%%%%%%%%%%%%%%%%
\begin{figure}
\centering
\resizebox{9cm}{!}{\includegraphics{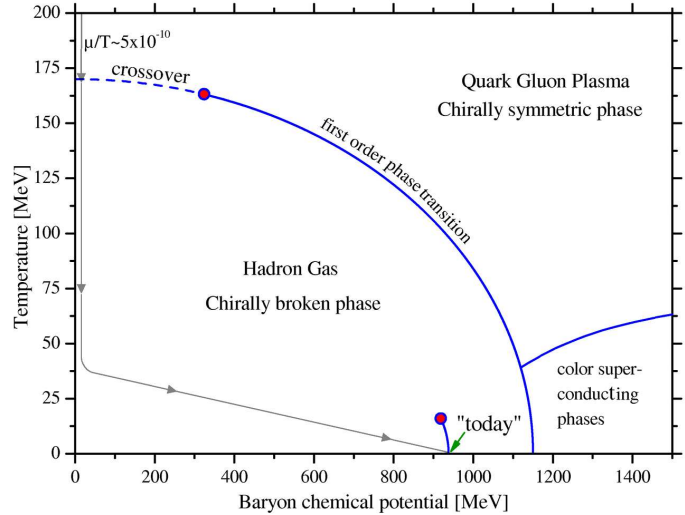}}	
\caption{Outline of a possible phase diagram for strongly interacting matter. In the conventional picture the early universe evolves at small baryo\-chemical potential, where lattice QCD predicts a crossover. The evolution of the universe follows the arrow, see also \cite{Fromerth02}. (Figure taken from Ref.~\refcite{Schettler11}.)} 
\label{figQCDPhaseDiagram} 
\end{figure}
%%%%%%%%%%%%%%%%%%%%%%%%%%%%%%%%%%%%%%%%%
%%%%%%%%%%
The proportionality holds for a radiative medium with $g(T)$ effective relativistic degrees of freedom. A GW starts to decay when the wavelength $\lambda$ becomes smaller than the Hubble radius $H^{-1}$, i.e. at horizon entry. $H^{-1}$ depends on $g$ and $T$ whereas, by contrast,  $\lambda$ grows proportional to the scale factor $a$---independently of the properties of the medium.
Thus, a drop in the relativistic degrees of freedom, as expected during the QCD phase transition, can modify the spectrum. In this contribution we present our results for the primordial GW spectrum after different types of QCD phase transitions including one at high $\mu_\mathrm B$ featuring a short period of inflation \cite{Boeckel10, Schettler11}. The little inflation also modifies the power spectrum of large-scale structure up to an enclosed dark matter mass of $M \sim 10^6 M_\odot$ and dilutes the cold dark matter density up to a factor of $10^{-9}$. This requires a reduced WIMP annihilation cross section or a larger WIMP mass which could be probed at the LHC. During a first order transition collisions of charged bubbles can generate magnetic fields which could be the seed fields for the galactic and extragalactic magnetic fields observed today.
We point out that the concept of a short inflationary period during the QCD phase transition has also been introduced earlier by K\"ampfer et al. \cite{Boyko90, Jenkovszky90} and Borghini et al. \cite{Borghini00}

%%%%%%%%%%%%%%%%%%%%%%%%%%%%%%%%%%%%%%%%%
\section{QCD Phase Transition with a Little Inflation}
%%%%%%%%%%%%%%%%%%%%%%%%%%%%%%%%%%%%%%%%%

\begin{figure}[t]
\centering
\resizebox{9cm}{!}{\includegraphics{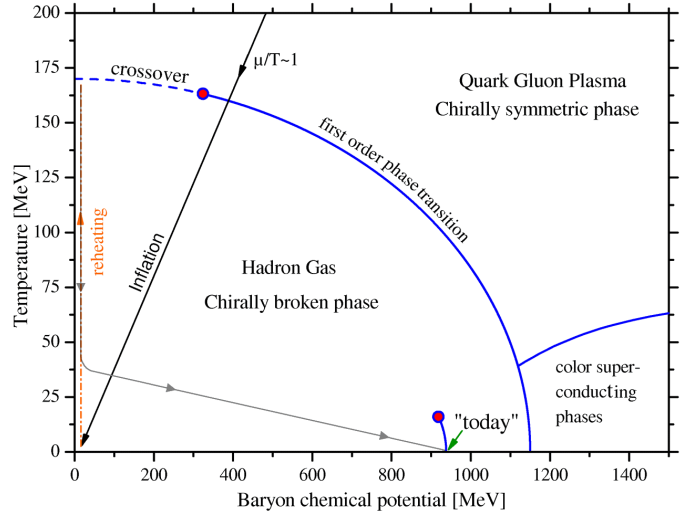}}
\caption{Possible track of the universe through the phase diagram for strongly interacting matter. The system crosses the first order line starting from high chemical potential $\mu_\mathrm B$. It is supercooled during a short period of inflation followed by the actual phase transition when reheating sets in. (Figure taken from Ref.~\refcite{Schettler11}.)}
\label{figInflQCDPhaseDiagram}
\end{figure}

From lattice calculations we know that for small baryo\-chemical potential strongly interacting matter undergoes a crossover from the hadronic phase to the quark-gluon plasma. As observations of the cosmic microwave background and measurements of the abundances of light elements suggest, the baryon to photon ratio in the early universe and equivalently the baryo\-chemical potential are well within the crossover region of the QCD phase diagram: $n_\mathrm B / s \sim n_\mathrm B / n_\gamma \sim \mu_\mathrm B / T \sim 10^{-9}$, see \fref{figQCDPhaseDiagram}. This ratio is conserved except at baryogenesis and during a first order phase transition with release of latent heat. 
At $\mu_\mathrm B / T \sim 1$ there are no reliable data from lattice QCD, and a first order phase transition is not excluded. As will be discussed in the following, in this case there may be observable signals from the transition. But how do we circumvent the bounds on $\mu_\mathrm B$ set by big bang nucleosynthesis (BBN) data? A possible scenario is the following (see also \fref{figInflQCDPhaseDiagram}): Before QCD time a strong baryon asymmetry is produced, e.g. by an Affleck-Dine baryogenesis. This can lead to $\mu_\mathrm B / T \sim 1$ and therefore allows for a first order transition during which the universe is trapped in a false metastable vacuum. The energy density is then dominated by the potential of this state. Since the latter is constant in time, it leads to an exponential growth of the scale factor and thus to an additional period of (de Sitter) inflation:
\[H^2 \propto \rho = \mathrm{const.} \quad \Longrightarrow \quad \frac{\dot a}{a} = \mathrm{const.}  \quad \Longrightarrow \quad a \propto \exp(H \cdot t).\]
After the inflationary period 
%%%%%%%%%%
%%%%%%%%%%%%%%%%%%%%%%%%%%%%%%%%%%%%%%%%%
%%%%%%%%%%%%%%%%%%%%%%%%%%%%%%%%%%%%%%%%%
%%%%%%%%%%
the baryo\-chemical potential should be at the value needed for BBN to produce the right amount of light elements. Tagging quantities before the inflation with a subscript i and those afterwards with f we can write down
%%%
\[ \left(\frac{\mu_\mathrm B }{T}\right)_\mathrm f  \approx \left(\frac{a_\mathrm i}{a_\mathrm f }\right)^3 \left( \frac{\mu_\mathrm B}{T}\right)_\mathrm i \overset{!}{\approx} 10^{-9}.\]
Hence, to provide the necessary dilution of $\mu_\mathrm B$ only $N = \ln \left(a_\mathrm f /a_\mathrm i \right) \sim \ln (10^3) \sim 7$ e-folds are sufficient. This is few in comparison to the corresponding value $N \sim 50$ in standard inflation. 
Nevertheless, the production of hadronic bubbles within the supercooled quark-gluon plasma must be suppressed in order to delay the phase transition. The corresponding nucleation rate depends on the difference in the free energy of the two phases and on the surface tension \cite{Csernai92}. For high densities this is discussed in Refs. \refcite{Palhares10, Mintz10}.
The attenuation of the fraction $\mu_\mathrm B / T$ is accompanied by a strong supercooling which passes into a period of reheating after the little inflation. Thereby the transition temperature $T_\mathrm c$ is reached again and the evolution of the universe returns to its conventional path. 

Finally we should mention that the scenario described above is only one possibility for a first order QCD phase transition in agreement with observation. Another one includes a large lepton asymmetry and is discussed in Ref.~\refcite{Schwarz09}.

%%%%%%%%%%%%%%%%%%%%%%%%%%%%%%%%%%%%%%%%%
\section{Tensor perturbations and the QCD trace anomaly}
%%%%%%%%%%%%%%%%%%%%%%%%%%%%%%%%%%%%%%%%%

\begin{figure}[t]
\centering
\resizebox{10cm}{!}{\includegraphics{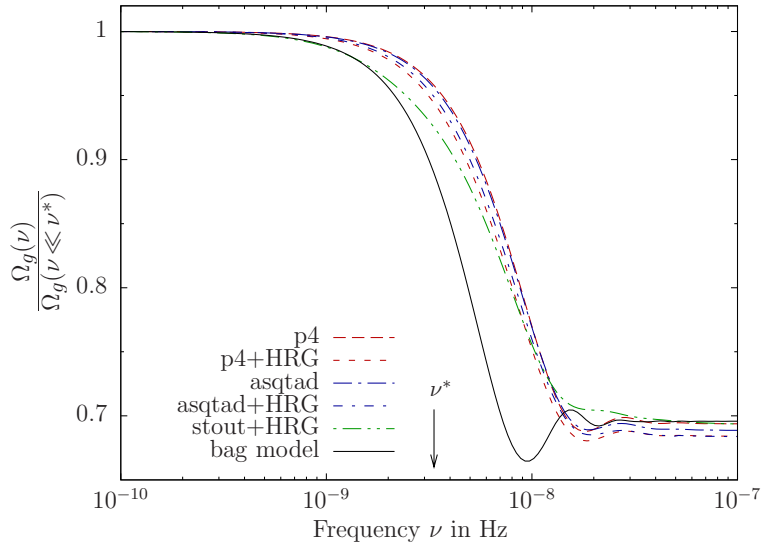}}
\caption{The energy density of GWs per logarithmic frequency interval after the QCD phase transition assuming entropy conservation. High frequency modes are suppressed by a factor of $\sim0.7$. The bag model calculation has been done with $T_\mathrm{c} = 180$~MeV.  The dashed curves represent calculations with lattice data. (Figure taken from Ref.~\refcite{Schettler11}.)}
\label{fig:SpectrumLattice}
\end{figure}

\begin{figure}[t]
\centering
\resizebox{10cm}{!}{\includegraphics{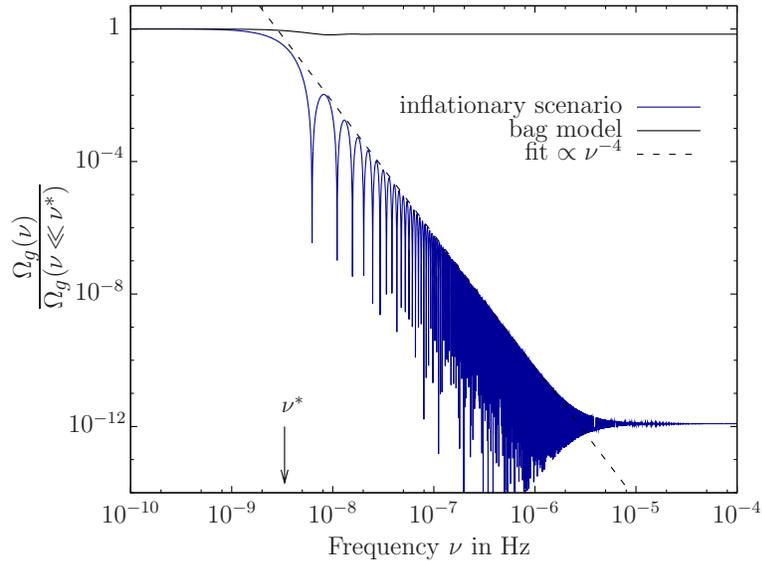}}
\caption{The GW spectrum after a QCD transition with a little inflation. The inflationary scenario entails a much stronger dilution than a transition with conserved entropy. (Figure taken from Ref.~\refcite{Schettler11}.)}
\label{fig:SpectrumInfl}
\end{figure}

In the future, observation of gravitational waves may allow for distinction between different types of QCD transitions. To see this we need the (gauge invariant) equation of motion for the tensor perturbation amplitude $v_k= a\cdot h_k$ in Fourier space. Making use of conformal time $\eta$ and denoting the corresponding derivatives with primes we get
\[ v''_k(\eta) + \left(k^2 - \frac{a''}{a}\right) v_k(\eta) = 0 \]
where 
\[ \frac{a''}{a} = \frac{4\pi G a^2}{3} \left(\rho-3p\right).\]
%%%%%%%%%%%%
%%%%%%%%%%%%%%%%%%%%%%%%%%%%%%%%%%%%%%%%%
\begin{figure}[t]
\centering
\resizebox{10cm}{!}{
\includegraphics{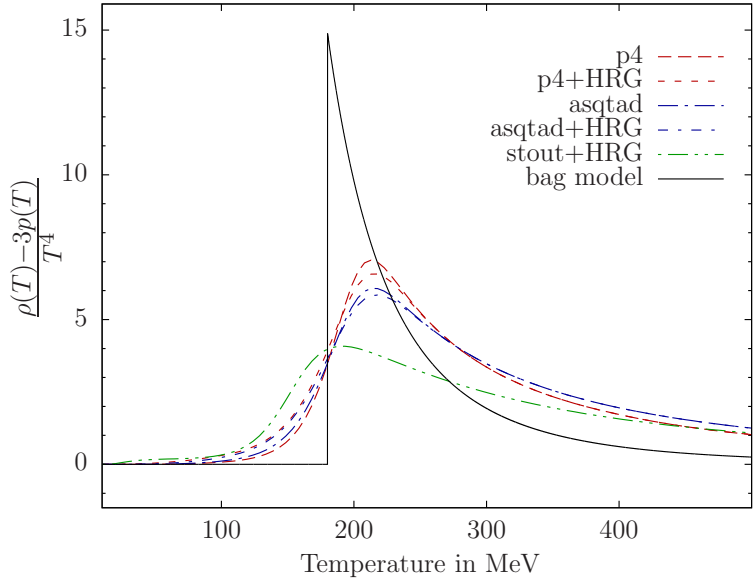}}
\caption{The trace anomaly of strongly interacting matter. The dashed curves depict parameterizations of results of lattice QCD. In part also the hadron resonance gas (HRG) has been taken into account. (Figure taken from Ref.~\refcite{Schettler11}.)}
\label{fig:TraceLattice}
\end{figure}
%%%%%%%%%%%%%%%%%%%%%%%%%%%%%%%%%%%%%%%%%
%%%%%%%%%%%%%
The last expression is proportional to the trace anomaly to which GWs are sensitive. In order to calculate the evolution of the primordial GW spectrum around QCD time we use parameterizations corresponding to several lattice calculations, see \fref{fig:TraceLattice}. The lattice results were obtained with improved staggered fermion actions (asqtad, p4, and stout) with and without including a hadron resonance gas (HRG) \cite{Bazavov09, Borsanyi10, Borsanyi10b}. The resulting GW spectra show a step-like structure at a critical frequency $\nu^*$ which corresponds to the mode that enters the Hubble horizon at the end of the phase transition. In \fref{fig:SpectrumLattice} the resulting energy density in GWs per logarithmic frequency interval is shown:
\[ \Omega_\mathrm g(k) = \frac{1}{\rho_\mathrm  c} \frac{d\rho_\mathrm g}{d\ln k} \quad\text{with $\rho_\mathrm c $ being the critical density.}\]
The spectra in \fref{fig:SpectrumLattice} are rather insensitive to details of the phase transition. This is because entropy is conserved during a phase transition described by the bag model or lattice data at vanishing $\mu_\mathrm B$. Entropy conservation in turn leads to
\[ \frac{\Omega_\mathrm  g(\nu \gg \nu^*)}{\Omega_\mathrm g(\nu \ll \nu^*)} = \left(\frac{g_\mathrm f}{g_\mathrm i}\right)^{1/3} \approx 0.7 \]
%%%%%%%%%%%%%%
%%%%%%%%%%%%%%%%%%%%%%%%%%%%%%%%%%%%%%%%%
%%%%%%%%%%%%%%%%%%%%%%%%%%%%%%%%%%%%%%%%%
%%%%%%%%%%%%%%
if we choose the typical values $g_\mathrm i = 51.25$ and $g_\mathrm f = 17.25$ as in Ref.~\refcite{Schwarz98}. However, in case of large entropy release, as it is expected at the end of an inflationary period, the resulting spectrum differs drastically: Because during inflation the successive entrance of modes into the horizon is reversed to a successive horizon exit, the high frequency modes in \fref{fig:SpectrumInfl} are strongly suppressed with respect to the ones with low frequency. In addition to the consequential drop $\Omega_\mathrm  g(\nu) \propto \nu^{-4}$ we observe strong oscillations within the spectrum. They emerge because each mode is at a different phase at horizon exit during inflation.
%%%%%%%%%%%%%
%%%%%%%%%%%%%%%%%%%%%%%%%%%%%%%%%%%%%%%%%
%%%%%%%%%%%%%%%%%%%%%%%%%%%%%%%%%%%%%%%%%
%%%%%%%%%%%%%

%%%%%%%%%%%%%%%%%%%%%%%%%%%%%%%%%%%%%%%%%
\section{Observations of gravitational waves}
%%%%%%%%%%%%%%%%%%%%%%%%%%%%%%%%%%%%%%%%%

\begin{figure}
\centering
\resizebox{9cm}{!}{\includegraphics{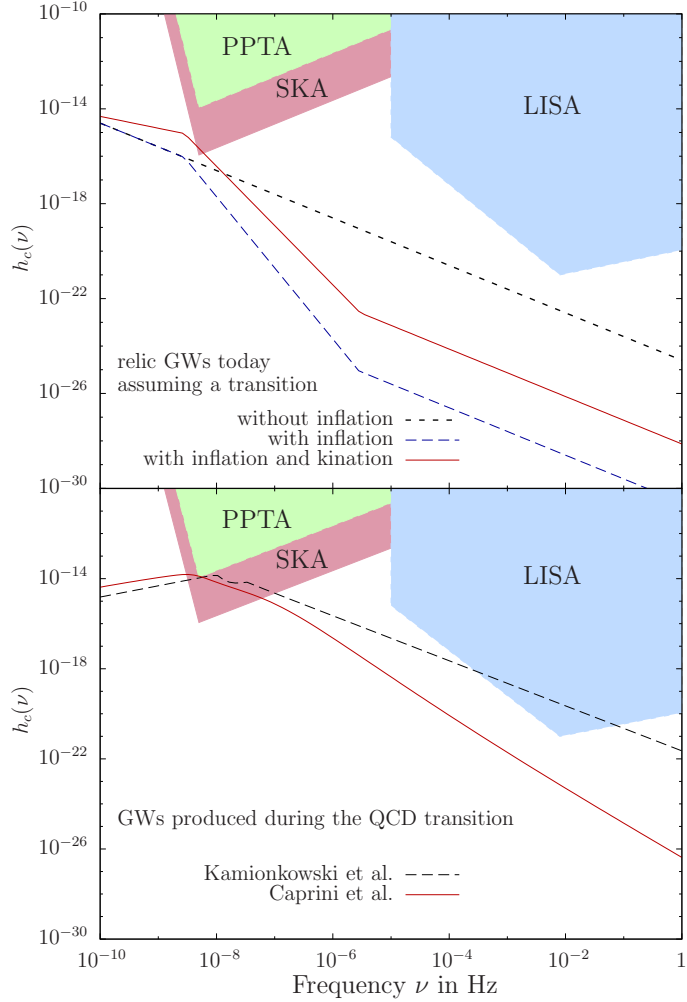}}
\caption{Different GW spectra after the QCD transition. The sensitivities of PPTA (Parkes Pulsar Timing Array), SKA (Square Kilometre Array) and LISA (Laser Interferometer Space Antenna) are displayed. PPTA already constrains the GW production during the QCD phase transition. The spectra in the lower panel are based upon analytic estimates in Ref.~\refcite{Kamionkowski94} (Kamionkowski et al.) and Ref.~\refcite{Caprini10} (Caprini et al.). See also Refs.~\refcite{Huber08, Kramer04, Kahniashvili08}. The displayed spectrum of relic GWs has the largest amplitude consistent with COBE\cite{Allen97}. (Figure taken from Ref.~\refcite{Schettler11}.)}
\label{figDetectors}	
\end{figure}

In \fref{figDetectors} different possible GW spectra after the QCD phase transition are displayed. 
In the upper panel the strain amplitude of the primordial GWs after a conventional QCD transition is compared with a spectrum resulting from the inflationary scenario (omitting the oscillations). 
The step frequency in the amplitude is close to the highest sensitivity of pulsar timing methods.
We also display a spectrum which is expected after a phase transition which additionally includes a long period of domination of kinetic energy of a scalar field (kination). 
Kination can occur during the process of reheating after inflation and here serves as an example of how future measurements of the GW spectrum could shed light on the time evolution of the equation of state in the early universe. 
The lower panel shows possible spectra of GWs emanating from bubble collisions and turbulences during a first order QCD transition. They most strikingly differ at high frequencies: In this regime the slopes of the spectra deviate because in Ref.~\refcite{Kamionkowski94} multi-bubble collisions are included. 
Note that these analytic estimates were done for a general first order transition at QCD time and do not include a short period of inflation.
For both we chose the maximal strength still being consistent with PPTA measurements. Recent analyses of data measured by the European Pulsar Timing Array EPTA \cite{vanHaasteren11} pose similar constraints on the spectra displayed in the lower panel. 
The Square Kilometre Array SKA is expected to push these limits further down. 
By contrast, a detection of the primordial GW spectrum seems impossible in the foreseeable future. 
However, comparison of Planck \cite{Planck05} measurements at $\nu \sim 10^{-18}$ Hz with those of the proposed Big Bang Observer BBO \cite{Corbin06} at $\nu \sim 1$ Hz will potentially allow for a discrimination between an inflationary transition and the standard one. 
%%%%%%%%%%%%%
%%%%%%%%%%%%%%%%%%%%%%%%%%%%%%%%%%%%%%%%%
%%%%%%%%%%%%%%%%%%%%%%%%%%%%%%%%%%%%%%%%%
%%%%%%%%%%%%%

%%%%%%%%%%%%%%%%%%%%%%%%%%%%%%%%%%%%%%%%%
\section{Summary}
%%%%%%%%%%%%%%%%%%%%%%%%%%%%%%%%%%%%%%%%%

We have presented an alternative scenario for the cosmological QCD phase transition which includes a short period of inflation. The scenario depends on a strong baryogenesis before the QCD scale allowing the universe to intersect the first order transition line which is suggested by effective models of QCD. The system needs to be trapped in a metastable vacuum state such that the corresponding vacuum energy leads to an inflationary phase.
Such a first order phase transition would generate signals which can potentially be observed: In this contribution we discussed the strong suppression of GWs with frequencies above $10^{-8}$ Hz and the production of GWs in bubble collisions and turbulences during the phase transition. We also mentioned effects on the dark matter power spectrum, different requirements on the WIMP annihilation cross section, and the production of primordial magnetic fields.

%%%%%%%%%%%%%%%%%%%%%%%%%%%%%%%%%%%%%%%%%
\section*{Acknowledgements}
%%%%%%%%%%%%%%%%%%%%%%%%%%%%%%%%%%%%%%%%%

One of us (JSB) is indebted to Walter Greiner for his continuing and stimulating interest in this work. This work is supported by BMBF under grant FKZ 06HD9127, by DFG within
the framework of the excellence initiative through the Heidelberg
Graduate School of Fundamental Physics, the International Max Planck
Research School for Precision Tests of Fundamental Symmetries
(IMPRS-PTFS), the Gesellschaft f\"ur Schwerionenforschung GSI
Darmstadt, the Helmholtz Graduate School for Heavy-Ion Research
(HGS-HIRe), the Graduate Program for Hadron and Ion Research (GP-HIR), and
the Helmholtz Alliance Program of the Helmholtz Association contract
HA-216 ``Extremes of Density and Temperature: Cosmic Matter in the
Laboratory''.

%%%%%%%%%%%%%%%%%%%%%%%%%%%%%%%%%%%%%%%%%

\end{document}